\documentclass[12pt,preprint]{aastex}

\usepackage{graphicx,epsfig,natbib,color}
\usepackage{lscape}
\usepackage{graphicx}
\usepackage{epsfig}
\usepackage{natbib}

\begin{document}
\title{Investigation of the Formation and Separation of An EUV Wave from the Expansion of A Coronal Mass Ejection}

\author{X. Cheng\altaffilmark{1,2,3}, J. Zhang\altaffilmark{2}, O. Olmedo\altaffilmark{4}, A. Vourlidas\altaffilmark{5}, M. D. Ding\altaffilmark{1,3}, Y. Liu\altaffilmark{6}}

\affil{$^1$ Department of Astronomy, Nanjing University, Nanjing 210093, China}\email{dmd@nju.edu.cn}
\affil{$^2$ School of Physics, Astronomy and Computational Sciences, George Mason University, Fairfax, VA 22030, USA}\email{jzhang7@gmu.edu}
\affil{$^3$ Key Laboratory for Modern Astronomy and Astrophysics (Nanjing University), Ministry of Education, Nanjing 210093, China}
\affil{$^4$ NRC, Naval Research Laboratory, Washington, DC 20375, USA}
\affil{$^5$ Space Science Division, Naval Research Laboratory, Washington, DC 20375, USA}
\affil{$^6$ Space Science Laboratory, University of California, Berkeley, CA 94720, USA}

\begin{abstract}

We address the nature of EUV waves through direct observations of the formation of a diffuse wave driven by the expansion of a coronal mass ejection (CME) and its subsequent separation from the CME front. The wave and the CME on 2011 June 7 were well observed by Atmospheric Imaging Assembly onboard Solar Dynamic Observatory. Following the solar eruption onset, marked by the beginning of the rapid increasing of the CME velocity and the X-ray flux of accompanying flare, the CME exhibits a strong lateral expansion. During this impulsive expansion phase, the expansion speed of the CME bubble increases from 100 km s$^{-1}$ to 450 km s$^{-1}$ in only six minutes. An important finding is that a diffuse wave front starts to separate from the front of the expanding bubble shortly after the lateral expansion slows down. Also a type-II burst is formed near the time of the separation. After the separation, two distinct fronts propagate with different kinematic properties. The diffuse front travels across the entire solar disk; while the sharp front rises up, forming the CME ejecta with the diffuse front ahead of it. These observations suggest that the previously termed EUV wave is a composite phenomenon and driven by the CME expansion. While the CME expansion is accelerating, the wave front is cospatial with the CME front, thus the two fronts are indiscernible. Following the end of the acceleration phase, the wave moves away from the CME front with gradually an increasing distance between them.

\end{abstract}
Online-only material: animations, color figures

\section{Introduction}

Coronal mass ejections (CMEs) are large-scale eruptive phenomena from the Sun. They can carry large amounts of plasma and magnetic field energy into the interplanetary space, which may have severe effects on space environment and human technological systems around the Earth \citep{gosling93,webb94}. A typical CME has a velocity of several hundred km s$^{-1}$, while the fastest one recorded is over 3000 km s$^{-1}$ \citep{Yashiro04}. The main acceleration of a CME usually occurs in the inner corona (e.g., $\leq$ 3.0 $R_\odot$) \citep{zhang01}. With the advent of the Atmospheric Imaging Assembly \citep[AIA;][]{lemen11} onboard Solar Dynamics Observatory (\emph{SDO}), details on the initiation and structural formation of CME starts to emerge. Recently, it was found that a twisted hot channel ($\sim$10 MK) as seen in AIA 131 {\AA} passband starts to form before the flare onset, and its subsequent rise results in the CME and the accompanying flare; this hot channel has been argued to be the existence of magnetic flux rope prior the eruption \citep{zhang11}. Following the eruption onset, a CME rises up impulsively with a strong acceleration, and also expands quickly along the lateral direction, producing a plasma bubble \citep{pat09b,cheng11}. The CME bubble is capable of driving a shock, which may generate the observable metric type II bursts \citep[e.g.,][]{liu09,ma11}.

One interesting phenomenon closely associated with CMEs is the globally-propagating bright feature in the corona dubbed the EIT or EUV wave since its discovery by \emph{SOHO} spacecraft \citep{thompson98,thompson99}. The physical nature of the moving front remains somewhat unclear. The front has been interpreted as a fast-mode magnetohydrodynamic (MHD) wave \citep{thompson99,wang00,warmuth01,veronig08,kienreich09,gopa09,pat09a,pat09b,liu11,olmedo11}, a soliton wave \citep{wills07}, or slow-mode wave \citep{wang09}. Whereas, others believe that it is not at all a true wave. \citet{chen02,chen05}, \citet{chen09}, and \citet{chen11} argued that the bright front results from the compression front driven by the successive stretching of the magnetic field overlying the erupting CME flux rope, although there should be a fast-mode wave ahead of the CME front. \citet{attrill07,attrill09} and \citet{dai10} claimed that the bright front is related to the magnetic reconnection between the outmost magnetic field of the CME and the magnetic loops from the quiet region. \citet{delannee08} suggested that the current shell in their numerical simulation can also form the bright front. In order to bring together these opposing views, some authors recently tend to appeal for the hybrid model combining both wave and non-wave explanations \citep[e.g.,][]{cohen09,liu10,downs11}. Details of various EUV wave models can be found in recent review papers by \citet{warmuth10}, \citet{gallagher11}, and \citet{zhukov11}.

In this Letter, we investigate the physical nature of the EUV wave. From the observations, we find that the wave front has two distinct evolution stages, i.e., a compression front forming the CME front in the early stage and a stand-alone wave front separating from the CME front in the later stage. The data used are mainly from AIA onboard \emph{SDO} and Sun Earth Connection Coronal and Heliospheric Investigation \citep[SECCHI;][]{howard08} onboard Solar TErrestrial RElations Observatory (\emph{STEREO}). Observations and results are presented in Section 2, followed by a summary and discussion in Section 3.

\section{Observations and Results}

\subsection{Overview of the CME Eruption}

\begin{figure} 
     \vspace{-0.0\textwidth}    
     \centerline{\hspace*{0.00\textwidth}
               \includegraphics[width=0.8\textwidth,clip=]{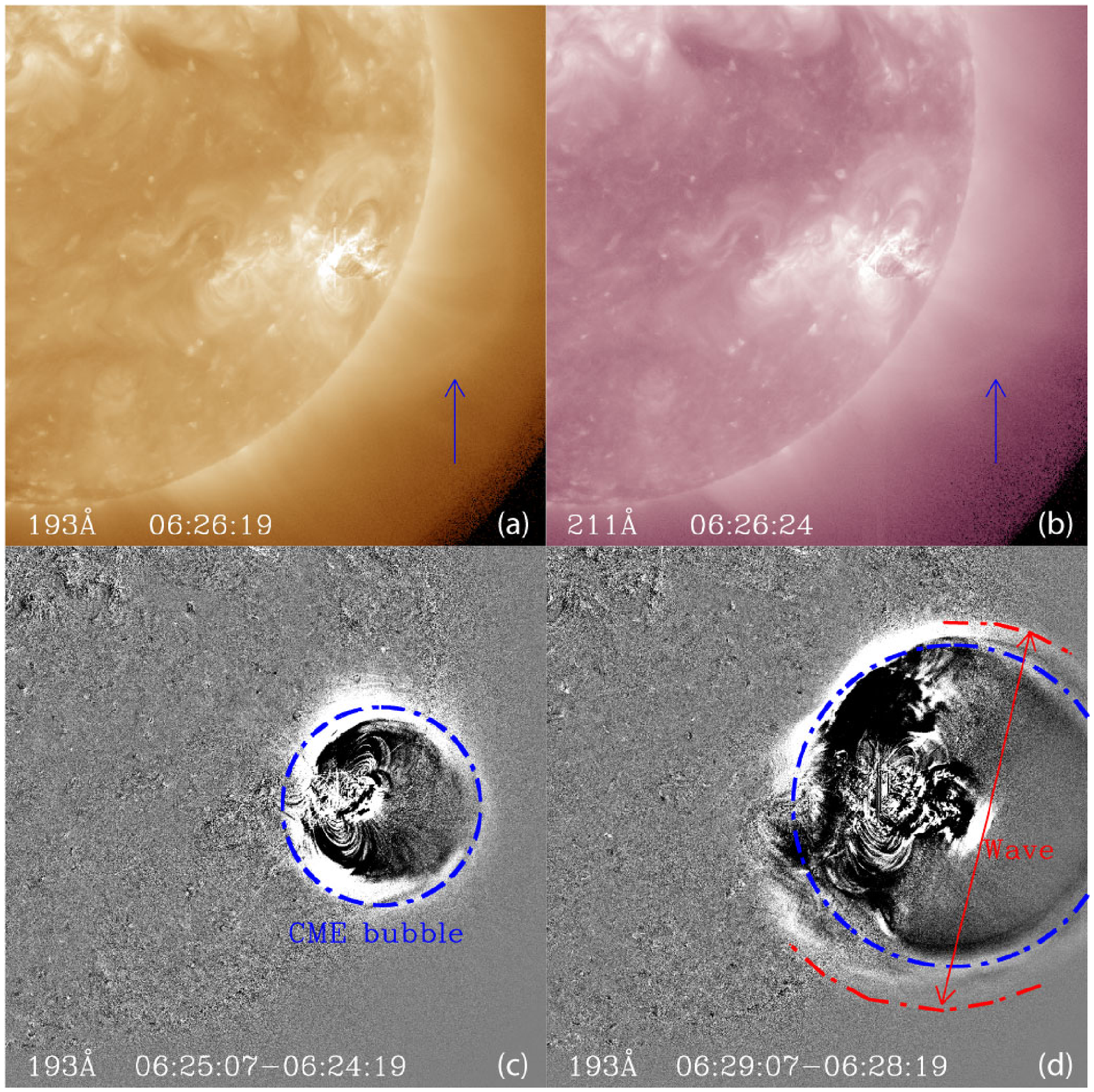}
               }
     \vspace{-0.001\textwidth}    
     \centerline{\hspace*{0.00\textwidth}
               \includegraphics[width=0.8\textwidth,clip=]{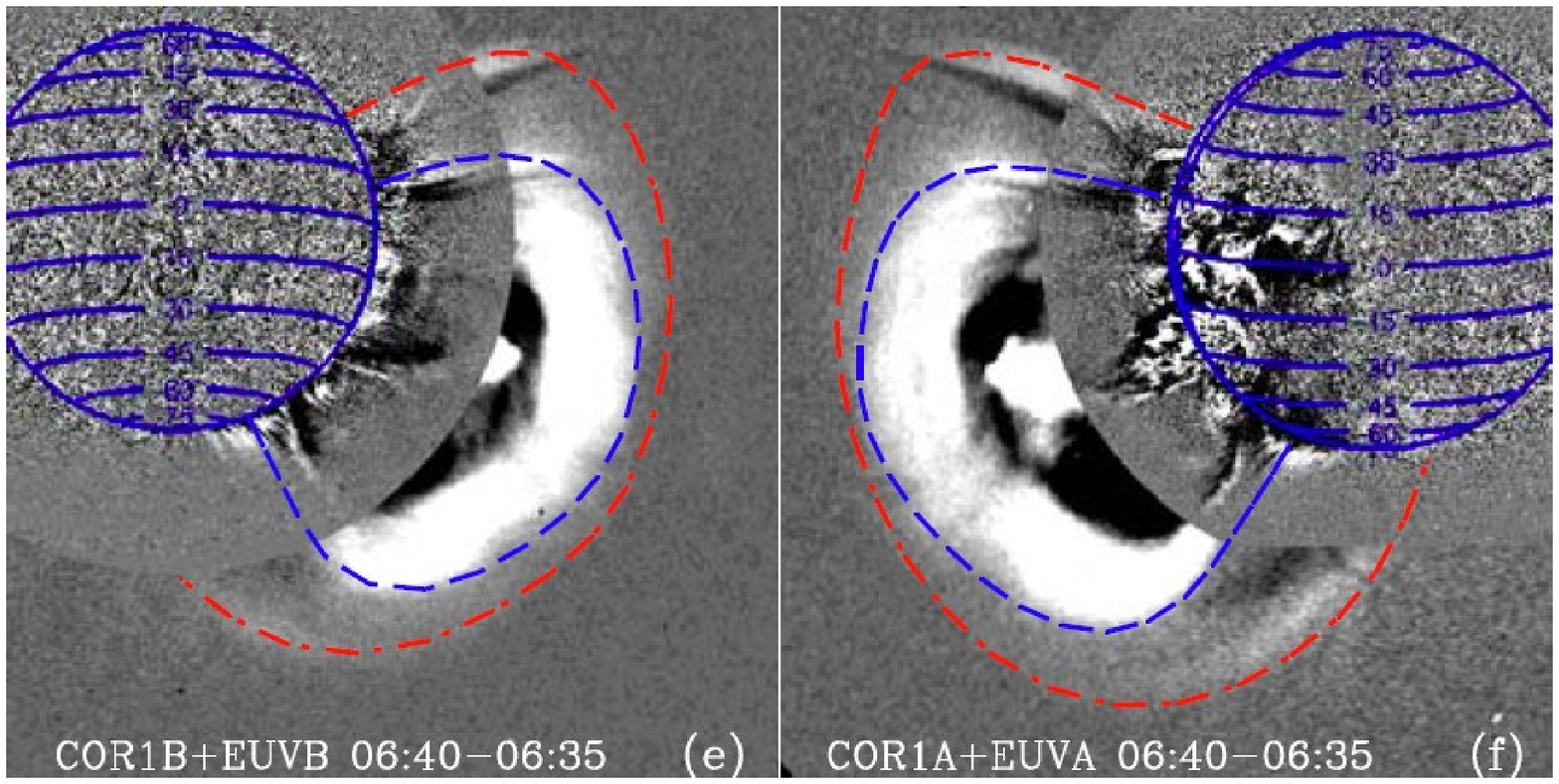}
               }

\vspace{0.0\textwidth}   
\caption{(a) and (b) AIA 193 {\AA} and 211 {\AA} images of the 2011 June 07 CME. The arrows indicate the CME bubble at 06:26 UT. (c) and (d) Sequence of 193 {\AA} running difference images showing the expanding and rising motion of the CME bubble. The blue dash-dotted lines denoted the circle fitting of the CME bubble; the red dash-dotted lines depict the diffuse wave front. (e) and (f) Difference images of the composite image of COR1 white light EUVI 195 {\AA}passband. The red and blue dash-dotted lines depict the diffuse wave front and the CME ejecta front, respectively. The blue solid lines indicate the heliographic latitude spaced by 15$^{\circ}$.} \label{f1}

(An animation of this figure is available in the online journal)
\end{figure}

On 2011 June 7, the active region (AR) NOAA 11226 produced an M2.5 class flare at the location of S22$^{\circ}$W53$^{\circ}$, which started at 06:16 UT and peaked at 06:30 UT. Following the onset of the flare, the overlying magnetic field of the AR expanded rapidly and formed a plasma bubble with a sharp front. At $\sim$06:26 UT, the bubble clearly appeared in AIA 193 {\AA} and 211 {\AA} images (Figure\ref{f1}(a) and (b)), best seen in the running difference images in Figure \ref{f1}(c). After $\sim$3 minutes, the bubble front started to leave the FOV of AIA (Figure \ref{f1}(c)). Subsequently, it developed into CME ejecta as seen in the FOV of COR1 (Figure \ref{f1}(e) and (f)).

\subsection{Rise and Early Expansion of the CME}

\begin{figure} 
     \vspace{-0.0\textwidth}    
     \centerline{\hspace*{0.00\textwidth}
               \includegraphics[width=0.6\textwidth,clip=]{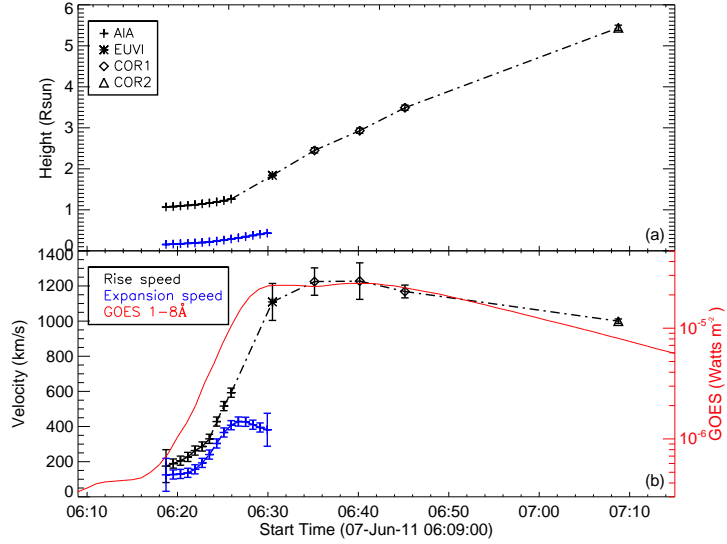}
               }

\vspace{0.0\textwidth}   
\caption{Kinematic evolution of the CME bubble. (a) Height-time plot of the CME bubble leading front (black line) and the radius-time plot of the bubble intrinsic radii from circular fitting (blue line). (b) The velocity evolution of the CME bubble leading front, and the intrinsic expansion velocity of the CME bubble (blue line). GOES Soft X-ray 1--8 {\AA} flux of the associated flare is also plotted (red line).} \label{f2}

\end{figure}

The shape of the CME bubble and ejecta can be clearly seen and tracked, thanks to the high cadence high quality AIA observations. So we are able to study the structural and kinematic evolution of the CME with high precision. The CME bubble is fitted as a circle in the AIA FOV, from which the radius of the bubble is obtained. The top of the circle is regarded as the height of the bubble front. Figure \ref{f1}(c) and (d) display the fitting result, in which the bubble is represented by the blue dash-dotted lines. Once the CME entered the FOV of COR1 (Figure \ref{f1}(e) and (f)) and COR2 (not shown here), it appeared very similar to a flux rope structure. Thus, we use the graduated cylindrical shell (GCS) model \citep{thernisien06,thernisien09} to model the 3D structure of the CME. The height-time variation of the CME front is shown in Figure \ref{f2}(a), along with the radius-time plot of the CME bubble in the AIA FOV.

Based on the height-time data, processed by the spline smoothing, we calculate the radial velocity of the CME front using a piece-wise numerical derivative method. The temporal evolution of the velocity is plotted in Figure \ref{f2}(b), in which we also plot the $GOES$ soft X-ray 1--8~\AA\ flux. One can see that the CME accelerated during the rise phase of the flare; the radial velocity of the CME increased from $\sim$100 km s$^{-1}$ at 06:19 UT to $\sim$1200 km s$^{-1}$ at 06:35 UT. The average acceleration during this period is $\sim$1130 m s$^{-2}$. Moreover, we calculate the expansion velocity of the CME bubble based on the radius-time data (blue line in Figure \ref{f2}(b)). It is found that the expansion of the bubble experienced a different kinematic evolution. In the first seven minutes, the expansion velocity of the bubble quickly increased from $\sim$100 km s$^{-1}$ to $\sim$450 km s$^{-1}$ with an average acceleration of $\sim$830 m s$^{-2}$. Whereas, after $\sim$06:26 UT, the expansion of the bubble started to slow down although its front was still undergoing acceleration. Note that the uncertainty in the velocity calculation is mainly from the error in the height measurement, which is estimated to be 4 pixels (1700 km, 9400 km, and 44000 km for AIA, EUVI, and COR1 and COR2 respectively).

\subsection{Separation of a Diffuse Wave from the CME Bubble}

One interesting finding from studying this event is that a diffuse wave is clearly seen to separate from the sharp bubble front. The separating wave front is mostly visible at the flank of the bubble in the FOV of AIA (Figure \ref{f1}(d)). The separation between the wave and the bubble is also clearly seen in the SECCHI observations (Figure \ref{f1}(e) and (f)). Inspecting the evolution of the wave and the bubble, we find that the diffuse wave front is always close to the bubble front at the top (or leading fronts along the radial direction). While, at the flanks of the bubble, the standoff distance between the two fronts gradually increases with time. At 06:40 UT, the diffuse front had propagated further away from the bubble front as shown in Figure \ref{f1}(e) and (f), from which we also find that the wave front above the limb coincides very well with that on the disk.

\begin{figure} 
\vspace{-0.12\textwidth}    
     \vspace{-0.0\textwidth}    
     \centerline{\hspace*{0.00\textwidth}
               \includegraphics[width=0.48\textwidth,clip=]{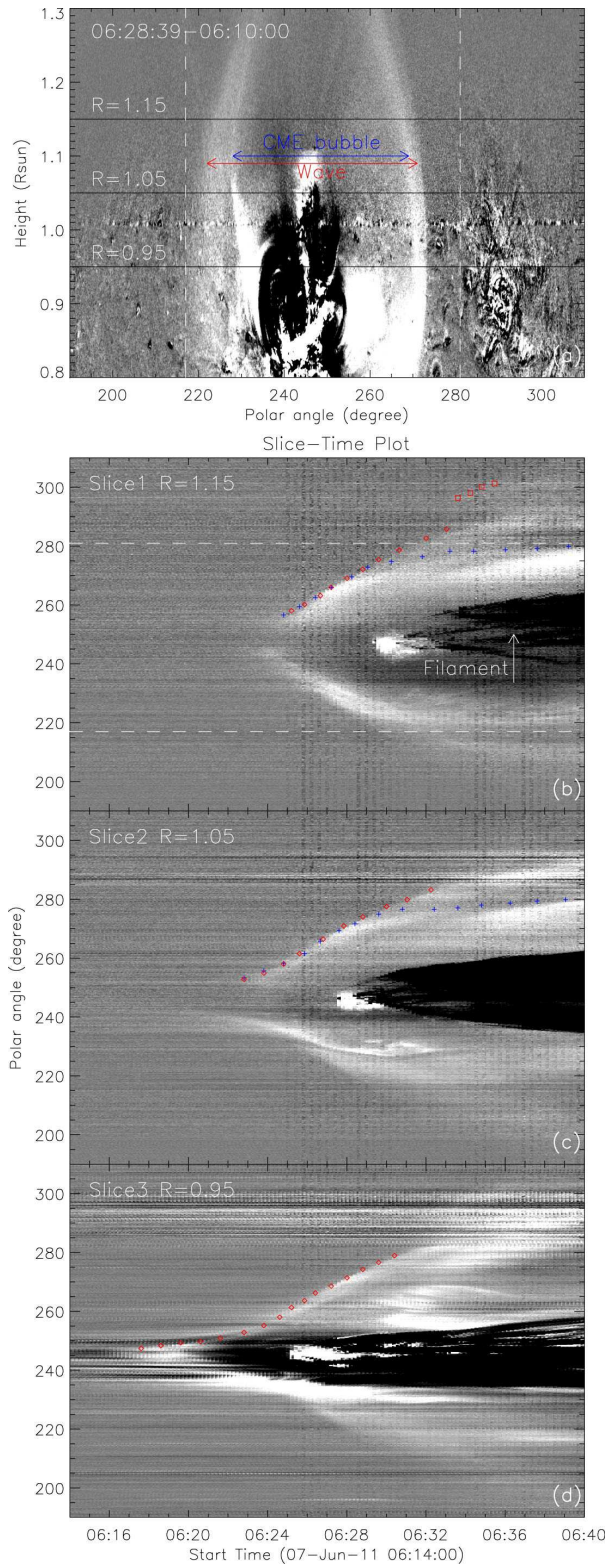}
               }

\vspace{-0.02\textwidth}   
\caption{(a) AIA 211 {\AA} difference image in the transformed helio-projective coordinate system. The red and blue arrow lines indicate the diffuse wave front and the CME bubble front, respectively. Two vertical dash lines indicate the boundaries of an AR in the north and the coronal hole in the south. The straight horizontal lines (black) show slices at the projected heliocentric distance of 1.15, 1.05 and 0.95$R_\odot$, respectively. (b--d) Slice-time plots showing the time evolutions of the bright fronts along the lateral direction at three different heights. The lateral displacement of the CME bubble front is indicated by the blue plus symbols, while the diffuse wave front is indicated by the red symbols. Two white horizontal dashed lines indicate the location of the same feature as indicated by white vertical dashed lines in (a).} \label{f3}

(An animation of this figure is available in the online journal)
\end{figure}

In order to investigate the detailed separation process of the wave front from the CME bubble, we transform the AIA images from the observed cartesian coordinates (x,y distance from sun center) into helio-projective coordinates (polar angle along the X-axis and projected heliocentric distance along the Y-axis). Figure \ref{f3}(a) shows the AIA 211 {\AA} difference images in the transformed system. Note that the 211 {\AA} difference images show the diffuse wave front best among all AIA passbands. One can see that the diffuse wave front can now be well distinguished from the sharp front of the CME bubble (see also the online movie for a better impression).

Next, from each transformed image we extract three horizontal slices located at heliocentric heights of 1.15, 1.05, and 0.95$R_\odot$, respectively. We then stack each slice vertically in a time sequence to make the slice-time plot. The results are shown in Figure \ref{f3}(b)-(d). From the stacking slice-time plots, one can see that the diffuse wave front has a different lateral evolution from the sharp bubble front. The wave front overlaps with the bubble front from $\sim$06:20 UT to $\sim$06:27 UT, i.e., the two fronts are exactly co-spatial, thus can not be discerned. After $\sim$06:27 UT, the wave front starts to separate from the bubble front, and the distance between the two fronts increases with time. The wave front continues to propagate, traversing through the AR in the north and the coronal hole in the south. On the other hand, the bubble's lateral expansion slows down significantly and stops near the AR in the north and the coronal hole in the south (dotted lines in Figure \ref{f3}(b) and (c) respectively). Note that there is only one front that can be clearly identified in the slice-time plot of the slice within the solar disk, because of the increased complexity of features on the disk.

\begin{figure} 
     \vspace{-0.0\textwidth}    
     \centerline{\hspace*{0.00\textwidth}
               \includegraphics[width=0.6\textwidth,clip=]{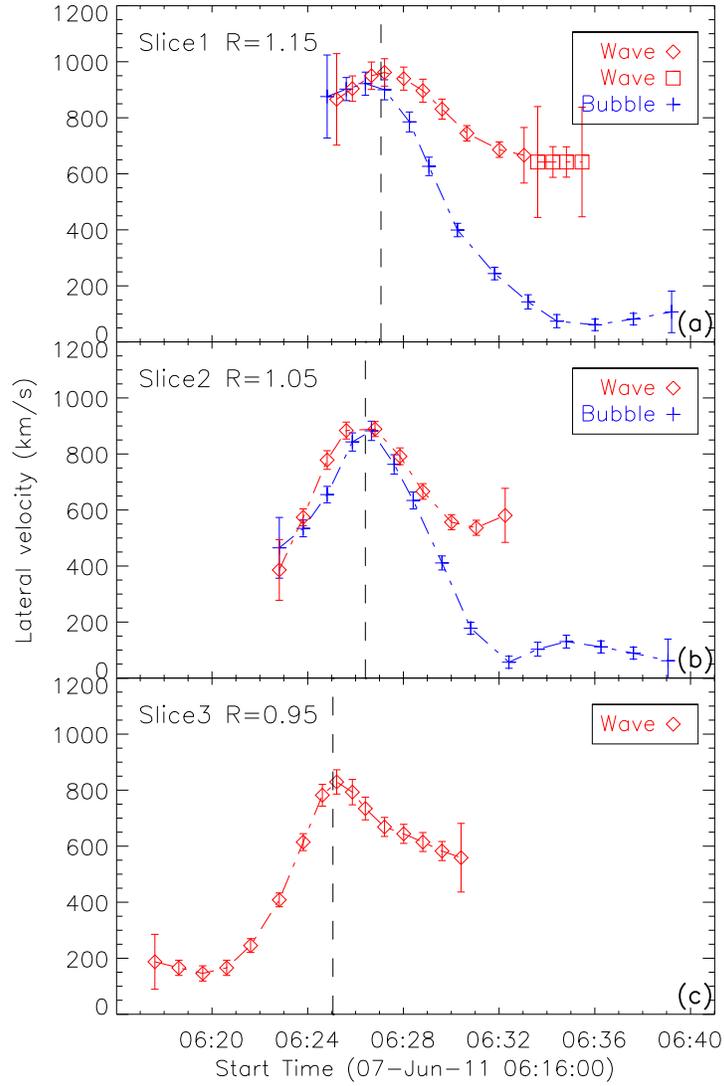}
               }

\vspace{0.0\textwidth}   
\caption{Kinematic evolution of the apparent lateral expansion of the CME bubble front (blue) and the diffuse wave front (red) along the line of the projected heliocentric distance (height) of 1.15 (a), 1.05 (b) and 0.95$R_\odot$ (c), respectively.} \label{f4}

\end{figure}

Through the slice-time plots, we can easily measure the lateral displacements of the CME bubble front and the wave front from a center position along the horizontal slice: the bubble front is denoted by the blue plus signs; and the wave front is shown by the red diamonds and squares (Figure \ref{f3}(b)-(d)). The resulting lateral propagation velocities are plotted in Figure \ref{f4}. The uncertainty in the velocity values results from the uncertainty of the lateral distance measurement, which is about 2800 km. It is worth noting that the lateral propagation velocity of the bubble along a same heliocentric distance (or a horizontal slice in the transformed helio-projective image) differs from its intrinsic expansion velocity; the apparent lateral velocity is a combination of the intrinsic expansion velocity and a geometric velocity caused by the rising motion of the bubble. Also note that, the apparent lateral velocity of the wave along a slice is close to its real propagation velocity because the slice is almost perpendicular to the wave front.

From Figure \ref{f4}(a), we can see that, at the heliocentric height of 1.15 $R_\odot$, the CME bubble front reached an apparent lateral velocity of 960 km s$^{-1}$ at $\sim$06:27 UT, which then quickly decreased to almost zero about 9 minutes later. For the wave front, it also accelerated to the lateral velocity of 960 km s$^{-1}$ at $\sim$06:27 UT, but the lateral velocity only decreased to $\sim$600 km s$^{-1}$ in the next nine minutes. Moreover, the wave traversed the nearby AR and continued to propagate at the velocity of $\sim$600 km s$^{-1}$ (squares in Figure \ref{f4}(a)) \citep[see also,][]{li11}. The apparent lateral propagation of the CME bubble and the wave at the heliocentric height of 1.05 $R_\odot$ is similar to that at 1.15 $R_\odot$. At $\sim$06:26 UT, both of them obtained the peak lateral velocity of $\sim$880 km s$^{-1}$. However, after $\sim$06:26 UT, the lateral velocity of the bubble front rapidly decreased, while the wave front only decreased to $\sim$600 km s$^{-1}$. Similarly, the wave front at the heliocentric height of 0.95 $R_\odot$ accelerated from 150 km s$^{-1}$ at $\sim$06:18 to 830 km s$^{-1}$ $\sim$06:25, and afterwards propagated freely with a velocity larger than $\sim$600 km s$^{-1}$.

Apparently, in the early evolution stage immediately following the eruption onset, the wave front can not be discerned from the CME bubble front, indicating that the wave front is still undergoing compression from the expanding bubble. The standoff distance between the two fronts is almost zero. Both of them obtain the maximum lateral velocity at the same time. When the CME bubble starts to decrease the velocity, the wave front starts to separate and propagate away from its driver. From Figure \ref{f4}, one can notice that the wave front has different peak lateral velocities at different heliocentric height.  The wave's lateral peak velocities increase with the heliocentric heights; the corresponding peak times of the velocities also delay with respect to the increasing height (Table \ref{tb}). Such increase of the peak lateral velocity and its time delay are most likely from the combination effect of the intrinsic expansion motion and the fast rising motion of the CME bubble.


\section{Summary and Discussion}

In this Letter, we focus on studying the separation process of two distinct fronts associated with a solar eruption that occurred on 2011 June 7. Following the eruption onset, the magnetic field of the source region quickly expands and forms a circular bubble. In the early stage of the eruption, the bubble expands strongly with an accelerating velocity. Afterwards, the apparent expansion velocity of the bubble close to the solar surface quickly decelerates to almost zero. In the meantime, a diffuse wave front starts to separate from the sharp bubble front. We conclude that the wave originates from the compression of the surrounding plasma by the impulsively expanding CME bubble. Due to a small standoff distance between the compression front and the driver front, the two fronts can not be distinguished during the early stage of the evolution when the driver is still undergoing acceleration. Through examining the radio data from CALLISTO radio spectrometer, we find that a type II radio burst started at $\sim$06:26 UT, almost at the same time (slightly earlier) as the wave started to separate from the driver. The radio observation suggests that a shock was generated at that time, indicating that the compression wave had just turned itself into a shock wave. Recent studies by \citet{pat09a} and \citet{pat10} also noticed two fronts: the wave front and the bubble front. However, due to the relative low cadence of the \emph{STEREO}-EUVI observations (5 minutes), the two distinct fronts were seen in only a few frames. Here, the AIA observations in every 12 seconds not only reveal the existence of two fronts but also the detailed separation process of the diffuse front from the sharp bubble front after the expansion of the bubble slows down. These observations clearly demonstrate that the separated diffuse front is a true MHD wave, driven by the early accelerating expansion of the CME bubble.

In conclusion, our observations help understand the physical nature of usually termed EUV waves. The EUV wave is actually a composite phenomenon, consisting of two distinct fronts, a wave front and a CME (compressed plasma) front. We further find that its evolution can be divided into two stages. The first stage takes place in the accelerating expansion phase of the CME bubble, which acts as the piston-driver of the MHD wave. During this stage, the wave front is coupled together with the compression front of the CME bubble. In the second stage, when the expanding velocity of the CME bubble slows down, the MHD wave front decouples from the compression bubble front, forms a distinct front, and propagates across the solar disk. The observational result for the coexistence of both wave and non-wave fronts are in general consistent with the models and numerical simulations for EUV waves by \citet{chen02,chen05}. We believe that the previous dispute about the nature of EUV waves resides in, at least partly, different parts of the composite phenomenon that the authors may have observed. In fact, the duration of the wave and CME compression front coupling is different for each event, depending on the dynamics of the CME and the surrounding environment.


\acknowledgements We thank P. F. Chen for many valuable comments that helped to improve the manuscript significantly. SDO is a mission of NASA's Living With a Star Program. X.C., and M.D.D. are supported by NSFC under grants 10673004, 10828306, and 10933003 and NKBRSF under grant 2011CB811402. X.C. is also supported by the scholarship granted by the China Scholarship Council (CSC) under file No. 2010619071. J.Z. is supported by NSF grant ATM-0748003 and NASA grant NNG05GG19G. A.V. is supported by NASA contract S-136361-Y.


\bibliographystyle{apj}


\begin{table*}
\caption{Properties of the apparent lateral expansion velocity of the CME bubble front and the diffuse wave front of 2011 June 07 solar eruption.}
\label{tb}
\begin{tabular}{ccc}
\\ \tableline \tableline
Height         & Peak velocity       & Time$^{a}$  \\
$(R_\odot)$    & (km s$^{-1}$)       & (UT)  \\
\tableline
1.15 &960$\pm$48 &06:27:03 \\
1.05 &880$\pm$27 &06:26:24 \\
0.95 &830$\pm$43 &06:25:03 \\
\tableline
\multicolumn{3}{p{6.5cm}}{${}^\mathrm{a}$ Time of the peak velocity.}\\
\end{tabular}
\end{table*}


%
%
\end{document}